\documentclass[twocolumn,showpacs,aps,amssymb,floatfix,prd,amsmath,preprintnumbers]{revtex4}
\usepackage{epstopdf}
\usepackage{capt-of}
\usepackage{graphicx}  
\usepackage{dcolumn}   
\usepackage{bm}
\begin{document}
\input epsf.tex
\title{Modelling wormholes in $f(R,T)$ gravity}

\author{
    P.H.R.S. Moraes$^{*}$, P.K. Sahoo$^{\dagger}$}
    \affiliation{$^{*}$ITA - Instituto Tecnol\'ogico de Aeron\'autica - Departamento de F\'isica, 12228-900, S\~ao Jos\'e dos Campos, S\~ao Paulo, Brazil \footnote{Email: moraes.phrs@gmail.com}}
    \affiliation{$^{\dagger}$Department of Mathematics, Birla Institute of Technology and Science-Pilani, \\ Hyderabad Campus, Hyderabad-500078, India \footnote{Email: pksahoo@hyderabad.bits-pilani.ac.in}}

\begin{abstract}

In this work we propose the modelling of static wormholes within the $f(R,T)$ extended theory of gravity perspective. We present some models of wormholes, which are constructed from different hypothesis for their matter content, i.e., different relations for their pressure components (radial and lateral) and different equations of state. The solutions obtained for the shape function of the wormholes obey the necessary metric conditions. They show a behaviour similar to those found in previous references about wormholes, which also happens to our solutions for the energy density of such objects. We also apply the energy conditions for the wormholes physical content. 

\end{abstract}

\pacs{04.50.kd.}


\keywords{$f(R,T)$ gravity; wormhole}

\maketitle


\section{Introduction}

Extended theories of gravity have been proposed aiming the explanation of some observable phenomena that hardly can be explained through General Relativity theory. As examples for some of these phenomena, we can quote dark energy \cite{riess/2004,peebles/2003}, dark matter \cite{aprile/2003,akerib/2017}, massive pulsars \cite{demorest/2010,antoniadis/2013}, super-Chandrasekhar white dwarfs \cite{howell/2006,das/2013}, among others, for which, gravity theories such as $f(R)$ \cite{de_felice/2010} and $f(\mathcal{T})$ \cite{yang/2011,cai/2016}, with $R$ and $\mathcal{T}$ being, respectively, the Ricci and torsion scalars, may propose some explanation. For a review on extended theories of gravity, we quote Reference \cite{capozziello/2011}.

Not only upon geometrical (or torsion) aspects extended theories of gravity are constructed, but also from the consideration of extra material contributions, as in the $f(R,\mathcal{L}_m)$ \cite{harko/2010} and $f(R,T)$ theories \cite{harko/2011}, with $\mathcal{L}_m$ being the matter lagrangian and $T$ the trace of the energy-momentum tensor. The material corrections are expected to come from the existence of imperfect fluids or quantum effects, such as particle production (check \cite{harko/2014,singh/2016}). 

Particularly, the $f(R,T)$ gravity theories have already shown to provide good alternatives for the issues quoted above, as it can be seen in References \cite{moraes/2015,mam/2016,sahoo/2015,zaregonbadi/2016,baffou/2015}, for example.

In the present article, we are concerned in modelling wormholes (WHs) in the $f(R,T)$ gravity. WHs are hypothetical passages which associate two different regions of the space-time and their matter content is described by an anisotropic energy-momentum tensor. 

They still have not been observed although attempts to achieve so were proposed \cite{tsukamoto/2016,kuhfittig/2014,rahaman/2014,zhou/2016,bambi/2013,tsukamoto/2017,li/2014,nandi/2006,harko/2009}. Anyhow, they were theoretically predicted a long time ago. A. Einstein and N. Rosen were the first to investigate WH solutions with event horizon \cite{einstein/1935}. M.S. Morris and K.S. Thorne many decades later have shown that WHs can be traversable if filled with exotic matter, which violates the energy conditions \cite{morris/1988}.

The motivation for working with WHs in the $f(R,T)$ gravity context can be seen as coming from the fact that WHs matter content is described by an imperfect anisotropic fluid while the $T-$dependence of the $f(R,T)$ theory, as mentioned above, may be related with the existence of imperfect fluids in the universe. Also, the quantum effects contained in $T$, or terms proportional to it, may be related to the mechanism of particle production, since those terms prevent the energy-momentum tensor of the theory to conserve. In this regard, the possibility of particle creation in WHs has already been analysed in  \cite{kim/1992,pan/2015}.

Our article is organized as follows: in Section II we present a review of the $f(R,T)$ gravity. In Section III, we present the WH metric and the conditions that must be satisfied by the shape function. In Section 4 we apply the WH metric in the $f(R,T)$ formalism. In Section 5 we construct different WH models from different hypothesis for their matter content. Our results are discussed in Section 6.

\section{$f(R,T)$ gravity}

The $f(R,T)$ theory of gravity starts from the following total action \cite{harko/2011}

\begin{equation}\label{e1}
S=\frac{1}{16\pi}\int d^{4}x\sqrt{-g}f(R,T)+\int d^{4}x\sqrt{-g}\mathcal{L}_m,
\end{equation}
where $f(R,T)$ is an arbitrary function of the Ricci scalar $R$ and trace of the energy-momentum tensor $T$, $g$ is the metric determinant and $\mathcal{L}_m$ is the matter Lagrangian density, related to the energy-momentum tensor as \cite{landau/1998}

\begin{equation}\label{e2}
T_{ij}= - \frac{2}{\sqrt{-g}}\left[\frac{\partial(\sqrt{-g}\mathcal{L}_{m})}{\partial g^{ij}}-\frac{\partial}{\partial x^{k}}\frac{\partial(\sqrt{-g}\mathcal{L}_m)}{\partial(\partial g^{ij}/\partial x^{k})}\right].
\end{equation}
Moreover, we will work with units such that $c=1=G$. 

Following the steps of Refs.\cite{harko/2010,harko/2011}, we assume $\mathcal{L}_m$ depends only on the metric components $g_{ij}$ and not on its derivatives, such that we obtain
\begin{equation}\label{e3}
T_{ij}= g_{ij}\mathcal{L}_{m} - 2\frac{\partial \mathcal{L}_{m}}{\partial g^{ij}}.
\end{equation}

By varying the action $S$ given in Eq.(\ref{e1}) with respect to the metric $g_{ij}$ provides the $f(R,T)$ field equations \cite{harko/2011}

\begin{multline}\label{e4}
f_R(R,T)\left(R_{ij}-\frac{1}{3} Rg_{ij}\right) + \frac{1}{6}f(R,T)g_{ij} \\=8\pi G \left(T_{ij}-\frac{1}{3}Tg_{ij}\right)-f_T(R,T)\left(T_{ij} -\frac{1}{3}Tg_{ij}\right)\\-f_T(R,T)\left(\theta_{ij}-\frac{1}{3}\theta g_{ij}\right)+\nabla_i\nabla_jf_R(R,T),
\end{multline}
with $f_R(R,T)=\partial f(R,T)/\partial R$, $f_T(R,T)=\partial f(R,T)/\partial T$ and
\begin{equation}\label{e5}
\theta_{ij}=g^{ij}\frac{\partial T_{ij}}{\partial g^{ij}}.
\end{equation}

The $f(R,T)$ (as well as the $f(R,\mathcal{L}_m)$) gravity theories are able to describe a non-minimal matter-geometry coupling in their formalism. This implies that in such theories, test particles moving in a gravitational field will not follow geodesic lines. The coupling between matter and geometry induces an extra force acting on the particles, which is perpendicular to the $4$-velocity.

It is interesting to note that the extra force depends on the form of the matter lagrangian \cite{harko/2014b}. It was shown in \cite{bertolami/2008} that by considering $\mathcal{L}_m=p$, with $p$ being the total pressure, the extra force vanishes. On the other hand, more natural forms for $\mathcal{L}_m$, such as $\mathcal{L}_m=-\rho$, with $\rho$ being the energy density, are more generic, in the sense that they do not imply the vanishing of the extra force.

In this paper, we assume the matter Lagrangian is $\mathcal{L}_m=-\rho$. Hence, Equation (\ref{e5}) can be written as
\begin{equation}\label{e6}
\theta_{ij}=-2T_{ij}-\rho g_{ij}.
\end{equation}

Let us assume the function $f(R,T)=R+2f(T)$, where $f(T)$ is an arbitrary function of $T$. The $f(R,T)$ gravity field equations (\ref{e4}) with (\ref{e6}) take the form

\begin{equation}\label{7}
R_{ij}-\frac{1}{2}Rg_{ij}=8\pi T_{ij}+2f^{\dagger}(T)T_{ij}+[2\rho f^{\dagger}(T)+f(T)]g_{ij},
\end{equation}
with $R_{ij}$ being the Ricci tensor and $f^{\dagger}(T)=df(T)/dT$. Assuming $f(T)=\lambda T$, with $\lambda$ being a constant, the above equations can be rewritten as

\begin{equation}\label{8}
G_{ij}=(8\pi+2\lambda) T_{ij}+\lambda(2\rho+T)g_{ij},
\end{equation}
with $G_{ij}$ being the usual Einstein tensor. Such an assumption for the $f(R,T)$ functional form was originally proposed by the $f(R,T)$ gravity authors themselves in \cite{harko/2011}. It has been applied to a high number of papers, such as \cite{moraes/2015,mam/2016,cm/2016,farasat_shamir/2015,das/2016,shabani/2017,das/2017,gu/2017}, among many others.

\section{Wormhole metric and its conditions}

The static spherically symmetric WH metric in Schwarzschild coordinates $(t,r,\theta, \phi)$ is \cite{morris/1988, visser/1995}
\begin{equation}\label{9}
ds^2=-U(r)dt^2+\frac{dr^2}{V}+r^2d\Omega^2,
\end{equation}
where $d\Omega^2=d\theta^2+\sin^2\theta d\phi^2$ and $V=1-b(r)/r$. The redshift function $U(r)$ and the shape function $b(r)$ in metric (\ref{9}) must obey the following conditions \cite{morris/1988, visser/1995}:

\begin{enumerate}
\item The radial coordinate $r$ lies between $r_0\leq r <\infty$, where $r_0$ is known as the throat radius.
\item At the throat $r=r_0$, $b(r)$ must obey the condition
\begin{equation}\label{10}
b(r_0)=r_0
\end{equation}
and for $r>r_0$, i.e., out of the throat,
\begin{equation}\label{11}
1-\frac{b(r)}{r}>0.
\end{equation}
\item $b(r)$ has to obey the flaring out condition at the throat, i.e.,
\begin{equation}\label{12}
b'(r_0)<1,
\end{equation}
with $'=d/dr$.
\item For asymptotically flatness of the spacetime
geometry, the limit
\begin{equation}\label{13}
\frac{b(r)}{r}\rightarrow 0 \ \ \ \text{as}\ \ \ \ \vert r\vert\rightarrow \infty
\end{equation}
is required.
\item $U(r)$ must be finite and non-vanishing at the throat $r_0$.
\end{enumerate}
One can consider {\bf $U(r)=constant$} to achieve the de Sitter and anti-de Sitter asymptotic behaviour. Since a constant redshift function can be absorbed in rescaled time coordinate, we consider $U(r)=1$, such as Refs.\cite{cataldo/2011,rahaman/2007}, among others.

\section{Field equations for wormholes in $f(R,T)$ gravity}

We consider an anisotropic fluid satisfying the matter content of the form
\begin{equation}\label{14}
T^i_j=\text{diag}(-\rho, p_r, p_l,p_l)
\end{equation}
where $\rho=\rho(r)$ is the energy density, $p_r=p_r(r)$ and $p_l=p_l(r)$ are respectively the radial and lateral (measured orthogonally to the radial direction) pressures. The trace $T$ of the energy momentum tensor (14) turns out to be $T=-\rho+p_r+2p_l$. 

The components of the field equations (8) for the metric (9) with (14) are 
\begin{equation}\label{15}
\frac{b'}{r^2}=(8\pi+\lambda)\rho-\lambda(p_r+2p_l),
\end{equation}
\begin{equation}\label{16}
-\frac{b}{r^3}= \lambda \rho+(8\pi+3\lambda)p_r+2\lambda p_l,
\end{equation}
\begin{equation}\label{17}
\frac{b-b'r}{2r^3}= \lambda \rho+\lambda p_r+(8\pi+4\lambda) p_l.
\end{equation}
The above set of field equations admits the solutions
\begin{equation}\label{18}
\rho=\frac{b'}{r^2(8\pi+2\lambda)},
\end{equation}
\begin{equation}\label{19}
p_r= - \frac{b}{r^3(8\pi+2\lambda)},
\end{equation}
\begin{equation}\label{20}
p_l= \frac{b-b'r}{2r^3(8\pi+2\lambda)}.
\end{equation}

\section{Models of wormholes}

In this Section, we will obtain models of WHs which are constructed from different hypothesis for their matter content.

\subsection{Model 1}

For our first model we assume the pressures $p_l$ and $p_r$ are related as
\begin{equation}\label{21}
p_l= np_r,
\end{equation}
where $n$ is an arbitrary constant. Such a relation was assumed in \cite{rahaman/2007,garcia/2010}, for instance.

Using (21) in (19) and (20), we can obtain
\begin{equation}\label{22}
b(r)=Ar^{1+2n},
\end{equation}
where $A$ is an integration constant.

Since the metric is asymptotically flat, Equation (22) above is consistent when $A$ is positive and $n$ is negative. This also satisfies the flaring out condition, i.e., $\frac{b-b'r}{b^2}>0$.

The shape function $b(r)$ is plotted versus $r$ in Figure 1 above for the values $A=1$ and $n=-0.4$. From such a figure, the fundamental WH condition, i.e., $b(r)<r$, is obeyed.

\begin{figure}[]
\includegraphics[scale=0.4]{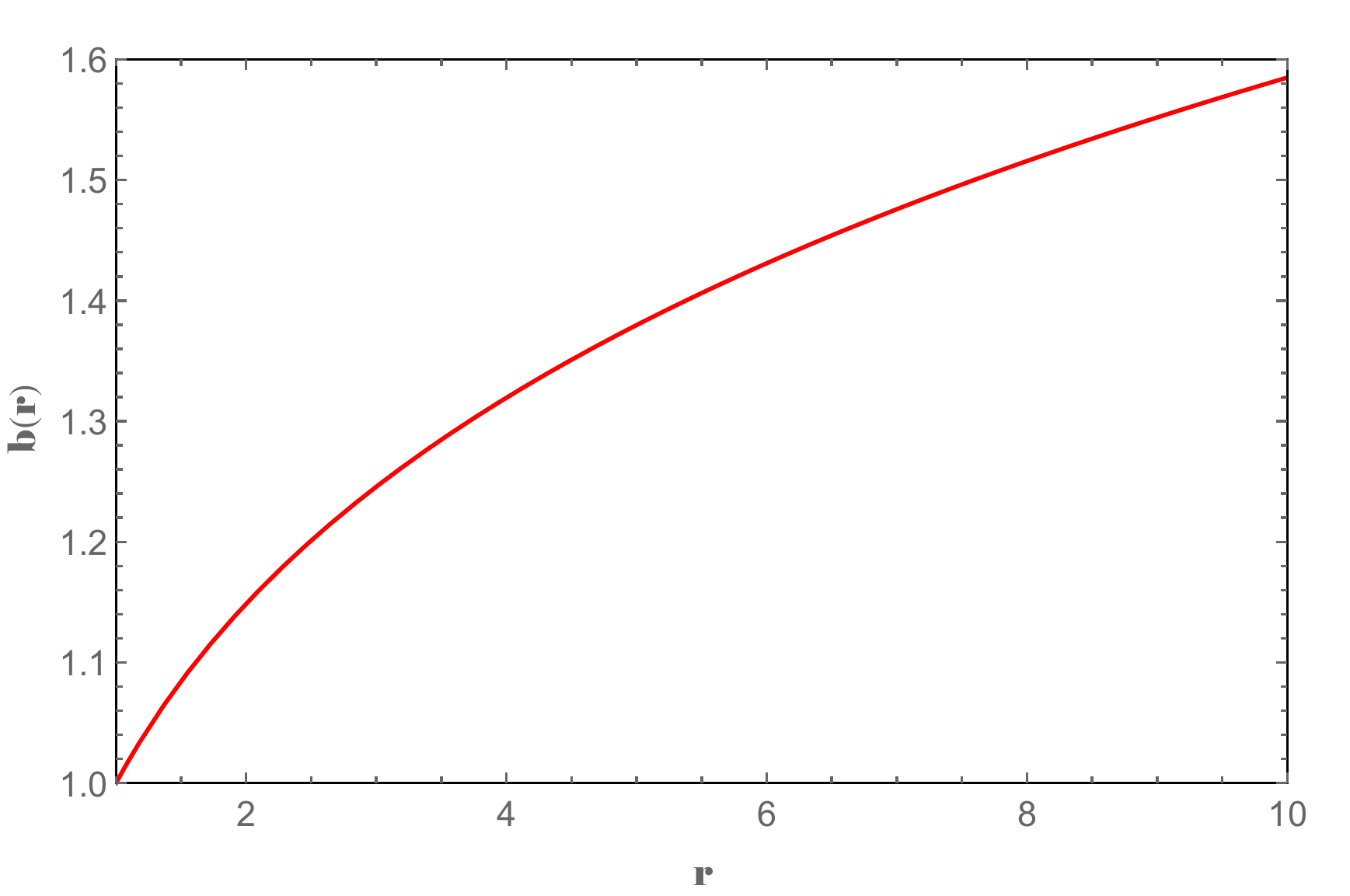}
\caption{\label{fig:1} The shape function $b(r)$, with $A=1$ and $n=-0.4$.}
\end{figure}

The throat of the WH occurs at $r=r_0$ so that Eq.(10) yields
\begin{equation}
A=\frac{1}{r_0^{2n}}.\label{23}
\end{equation}

The energy density will be
\begin{equation}\label{24}
\rho=\frac{A(1+2n)}{8\pi+2\lambda}r^{2(n-1)},
\end{equation}
while the radial and lateral pressures are
\begin{equation}\label{25}
p_r=\frac{-A}{8\pi+2\lambda}r^{2(n-1)}
\end{equation}
and
\begin{equation}\label{26}
p_l=\frac{-nA}{8\pi+2\lambda}r^{2(n-1)}.
\end{equation}
From Equations (24)-(26), we have
\begin{equation}\label{27}
\rho+p_r=\frac{nA}{4\pi+\lambda}r^{2(n-1)},
\end{equation}
\begin{equation}\label{28}
\rho+p_l=\frac{A(1+n)}{8\pi+2\lambda}r^{2(n-1)}.
\end{equation}

The existence of exotic matter in WHs violates the energy conditions. Specifically, the energy-momentum tensor violates the null energy condition (NEC) at the WH throat \cite{Hochberg/1998, DHochberg/1998}. One can observe from Eq.(27) the violation of NEC, i.e., $\rho+p_r\leq0$, which implies $\lambda>-4\pi$. NEC is presented in Fig.2 below. The validity region of $\rho+p_l\geq0$ is presented in Fig.3. 

\begin{figure}[!h]
\includegraphics[scale=0.4]{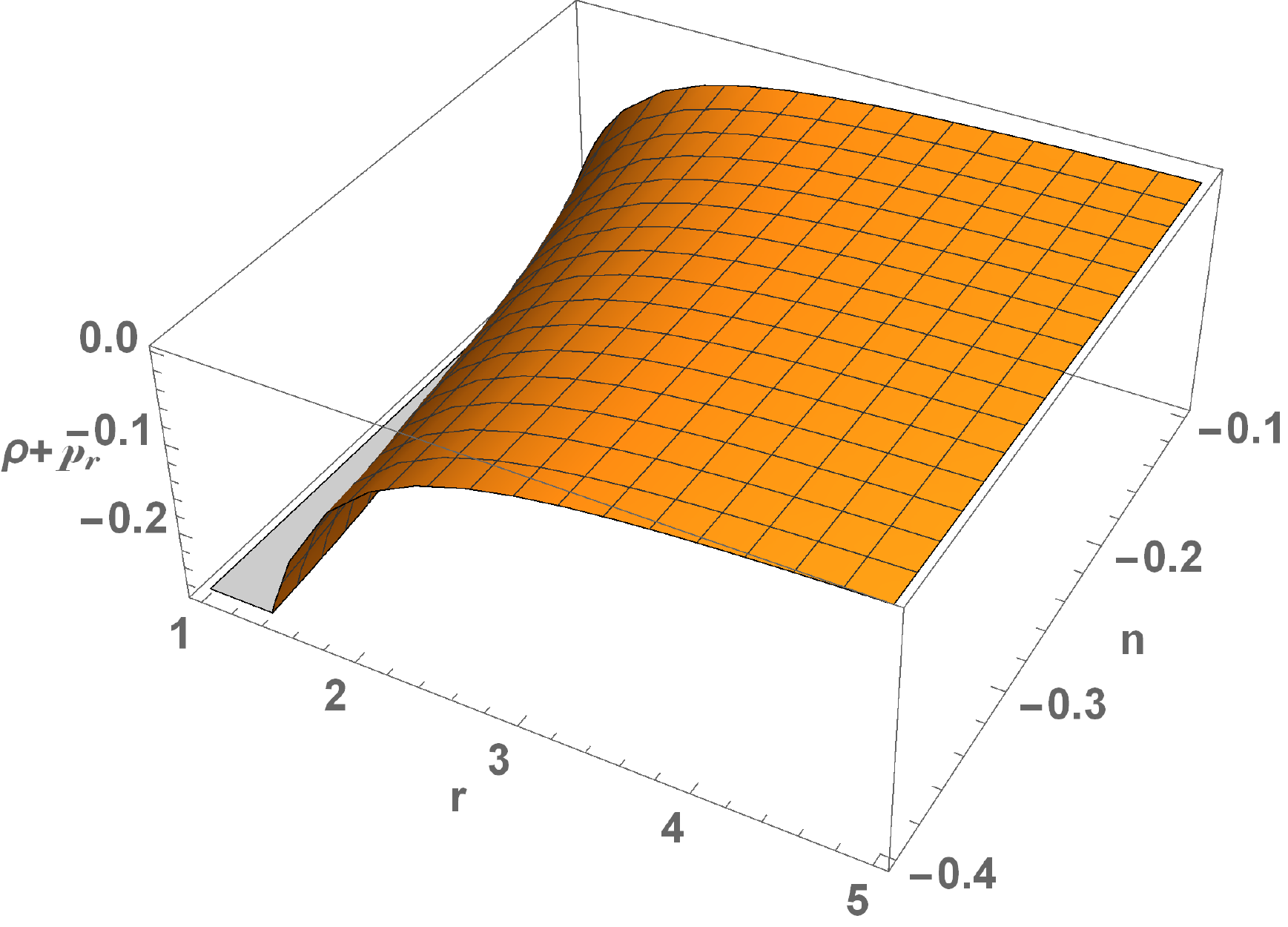}
\caption{\label{fig:2} Violation of NEC, $\rho+p_r\leq0$, with $A=1$ and $\lambda=-12$.}
\end{figure}

\begin{figure}[!h]
\includegraphics[scale=0.4]{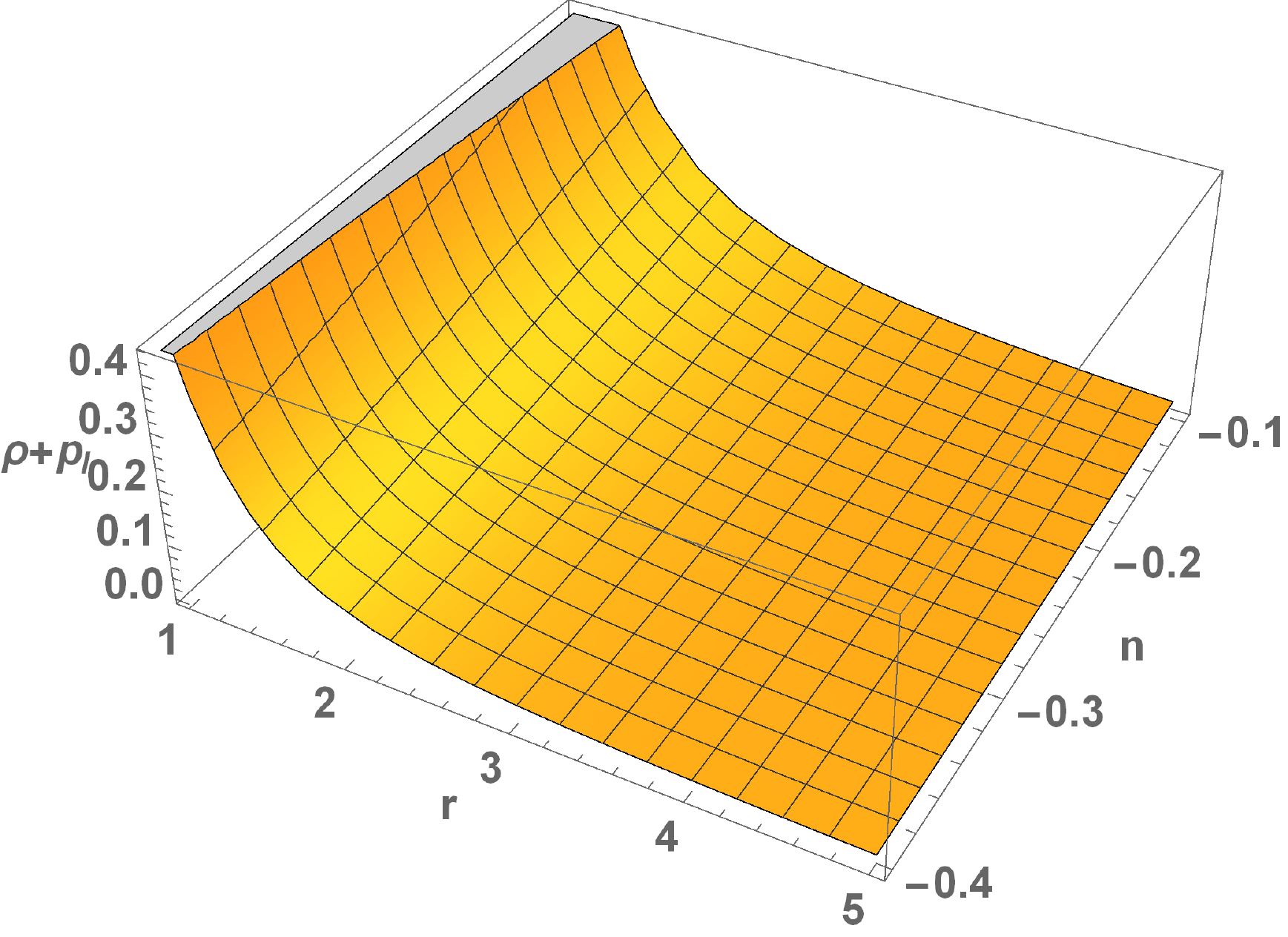}
\caption{\label{fig:3} NEC, $\rho+p_l\geq0$, with $A=1$ and $\lambda=-12$.}
\end{figure}

The dominant energy condition (DEC), $\rho\geq\vert p_r\vert$ and $\rho\geq\vert p_l\vert$, for this model is obtained from

\begin{equation}
\rho-p_r=\frac{A(1+n)}{4\pi+\lambda}r^{2(n-1)},\label{29}
\end{equation}
\begin{equation}
\rho-p_l=\frac{A(1+3n)}{8\pi+2\lambda}r^{2(n-1)}.\label{30}
\end{equation}
From Eqs.(29)-(30), DEC is plotted in Figs.4-5. 

\begin{figure}[!h]
\includegraphics[scale=0.4]{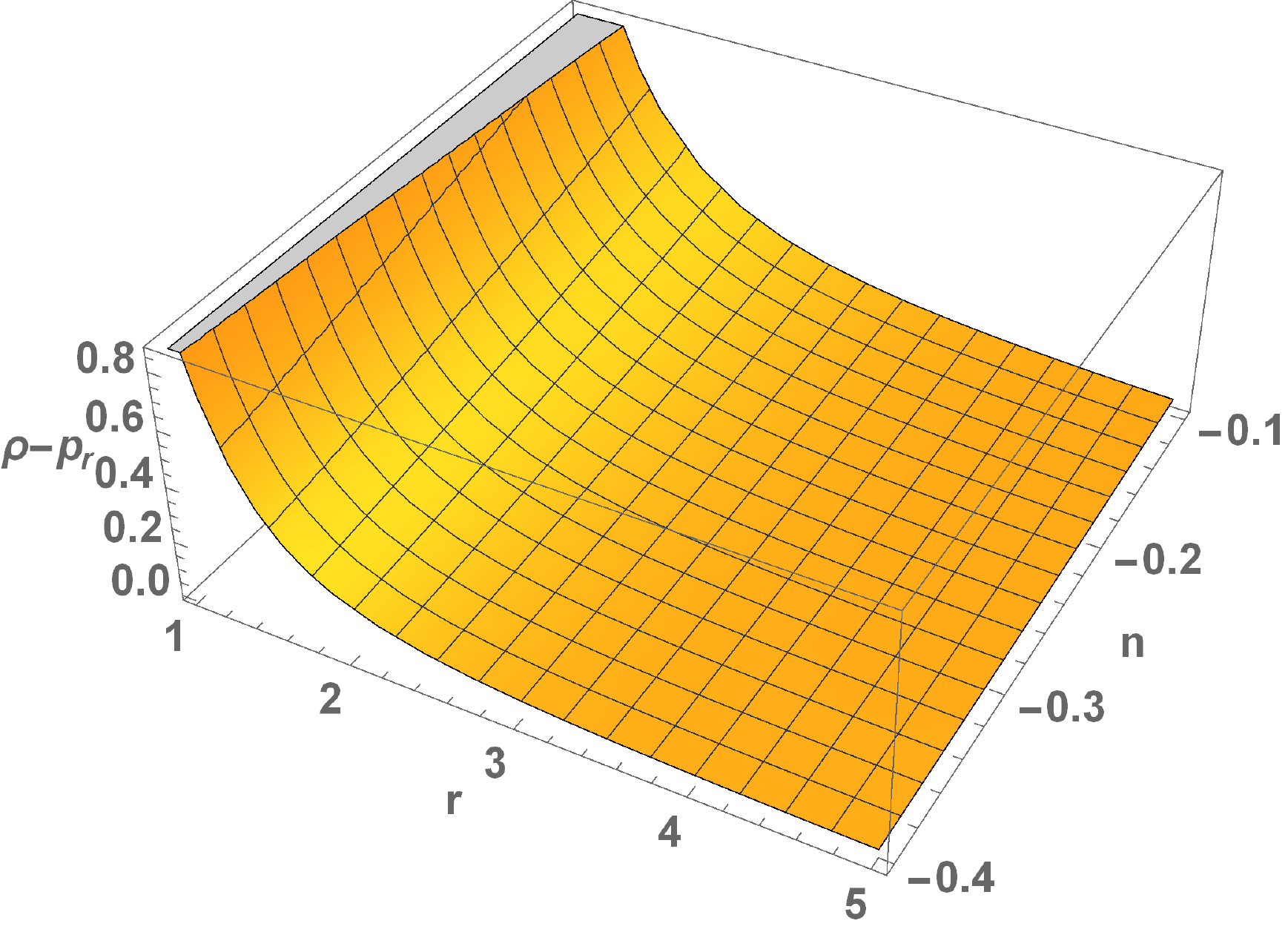}
\caption{\label{fig:4} DEC, $\rho\geq\vert p_r\vert$, with $A=1$ and $\lambda=-12$.}
\end{figure}

\begin{figure}[!h]
\includegraphics[scale=0.4]{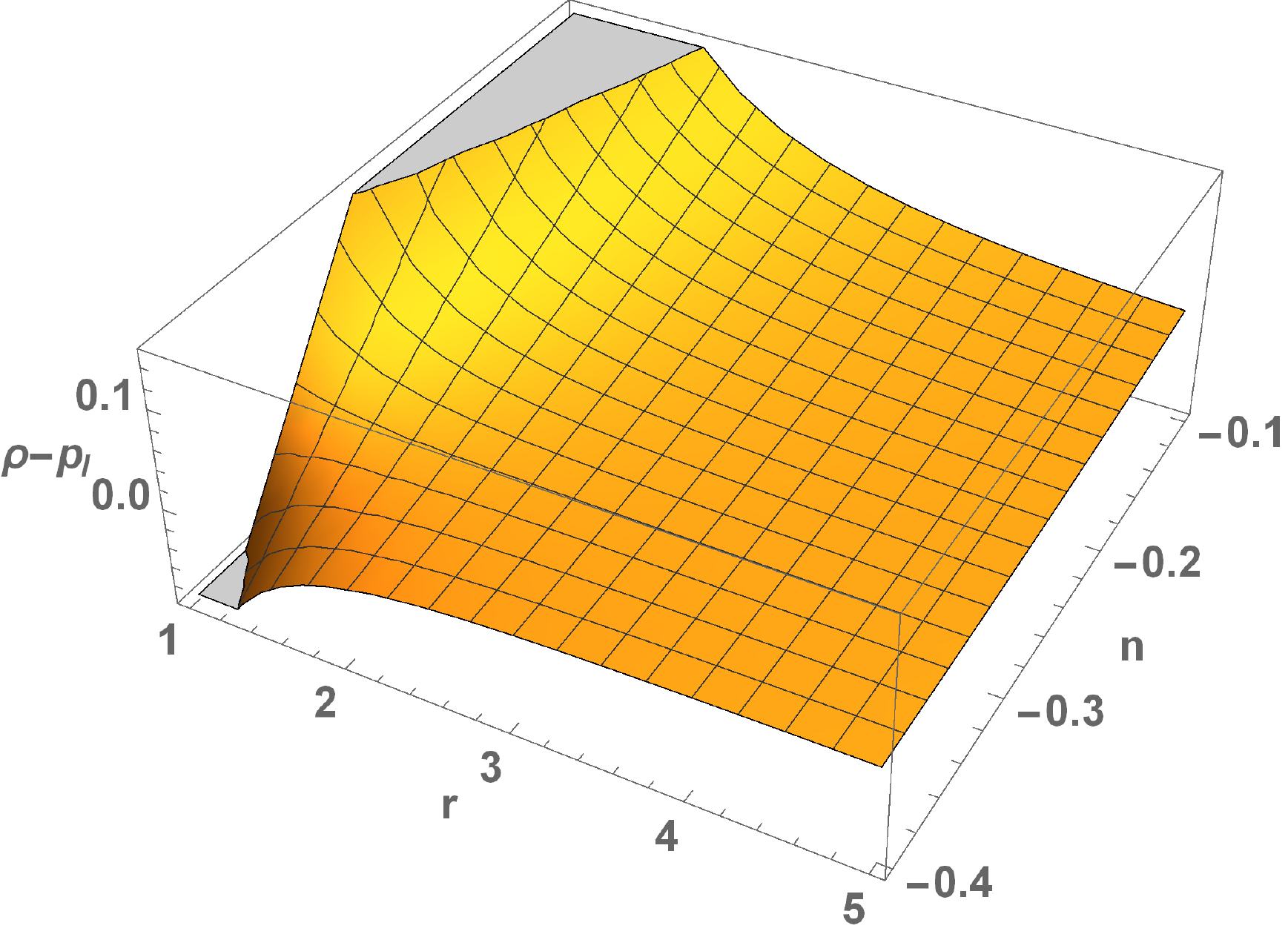}
\caption{\label{fig:5} DEC, $\rho\geq\vert p_l\vert$, with $A=1$ and $\lambda=-12$.}
\end{figure}

From Eqs.(18)-(20), the strong energy condition (SEC) yields $\rho+p_r+2p_l=0$.

\subsection{Model 2}

Let us consider that matter with equation of state (EoS)
\begin{equation}\label{31}
p_r+\omega(r) \rho=0
\end{equation}
is filling the WH, for which $\omega(r)$ is a positive function of the radial coordinate. The same EoS with varying parameter $\omega(r)$ was assumed in Reference \cite{rahaman/2009} by F. Rahaman and collaborators, for example.

Taking (31) into account, from Equations (18) and (19) we are able to obtain
\begin{equation}\label{32}
\omega(r)=\frac{b}{rb'}.
\end{equation}

Next, we will check two cases for Eq.(32) above, as it was previously made in \cite{rahaman/2009}.
\\

\textbf{Case I - $\omega(r)=\omega=\text{constant}$} \\

Let us take, as a first case, $\omega(r)=\omega=\text{constant}$. This implies, from Eq.(32), that
\begin{equation}\label{33}
b=b_0r^{1/\omega},
\end{equation}
where $b_0$ is an integrating constant. Since the metric is asymptotically flat, Equation (33) is consistent when $\omega>1$.

Using some particular values of the parameters, we plot $b(r)$ in Fig.6 below. It shows that for $r>r_0$, $b(r)-r<0$, an essential condition for the shape function to obey. $b(r)-r$ is also a decreasing function for $r>r_0$, which satisfies the flaring out condition $b'(r_0)<1$.

\begin{figure}[!h]
\includegraphics[scale=0.4]{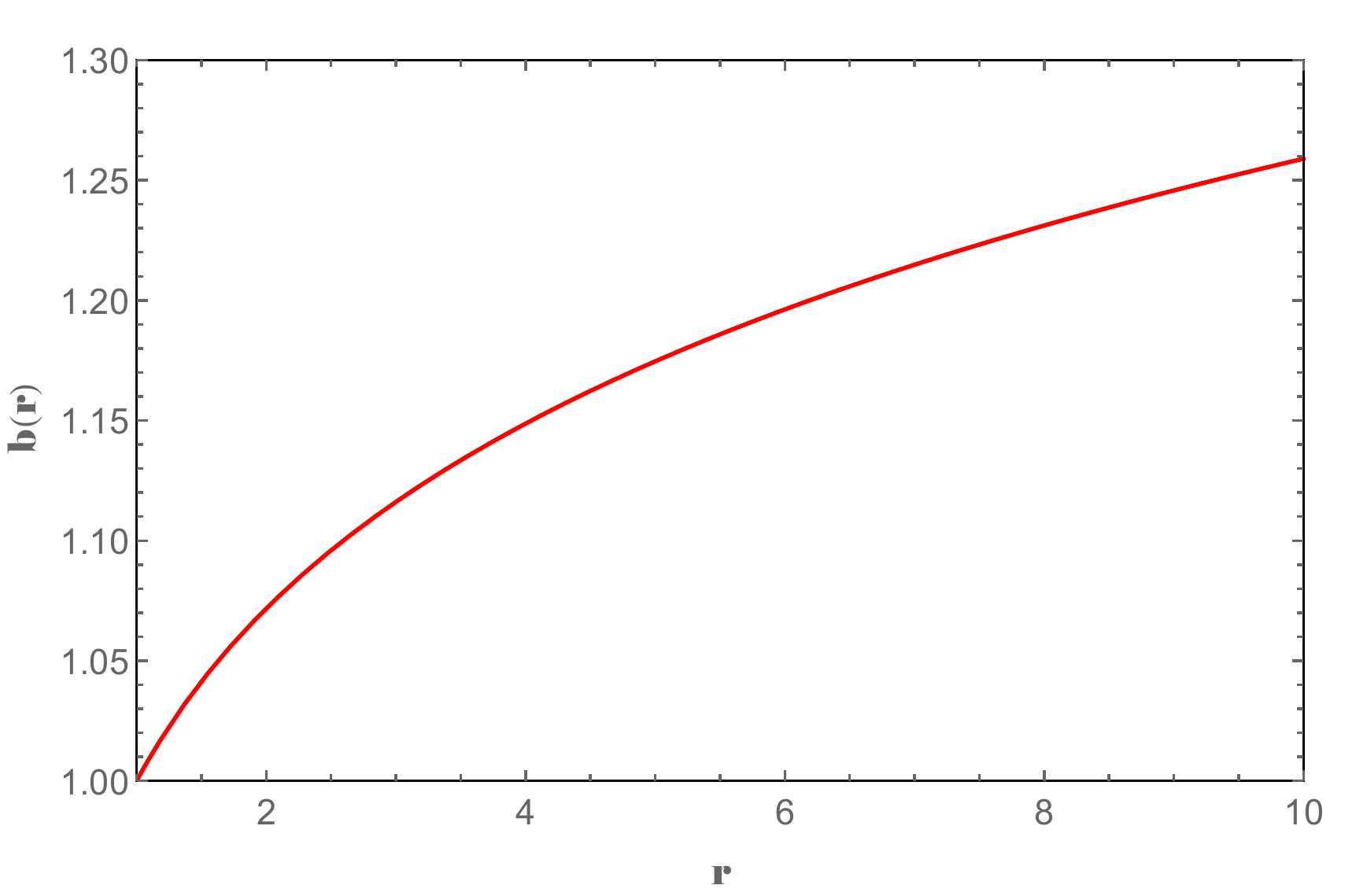}
\caption{\label{fig:6} $b(r)$ with $b_0=1$ and $\omega=10$.}
\end{figure}

The throat of the WH occurs at

\begin{equation}\label{34}
r_0=b_0^{\frac{\omega}{\omega-1}}.
\end{equation}


Using Eq.(33) in Eqs.(18)-(20), we have
\begin{equation}\label{35}
\rho=\frac{b_0r^{\frac{1-3\omega}{\omega}}}{\omega(8\pi+2\lambda)},
\end{equation}
\begin{equation}\label{36}
p_r=-\frac{b_0r^{\frac{1-3\omega}{\omega}}}{8\pi+2\lambda},
\end{equation}
\begin{equation}\label{37}
p_l=\frac{b_0(\omega-1)r^{\frac{1-3\omega}{\omega}}}{2\omega(8\pi+2\lambda)}.
\end{equation}

NEC for this model is obtained as
\begin{equation}\label{38}
\rho+p_r=\frac{b_0(1-\omega)r^{\frac{1-3\omega}{\omega}}}{\omega(8\pi+2\lambda)},
\end{equation}

\begin{equation}\label{39}
\rho+p_l=\frac{b_0(1+\omega)r^{\frac{1-3\omega}{\omega}}}{2\omega(8\pi+2\lambda)}.
\end{equation}

From (38) the violation of NEC happens for $b_0>0$ and $\lambda>-4\pi$. NEC for this model is presented in Figs.7-8 below.

\begin{figure}[!h]
\includegraphics[scale=0.4]{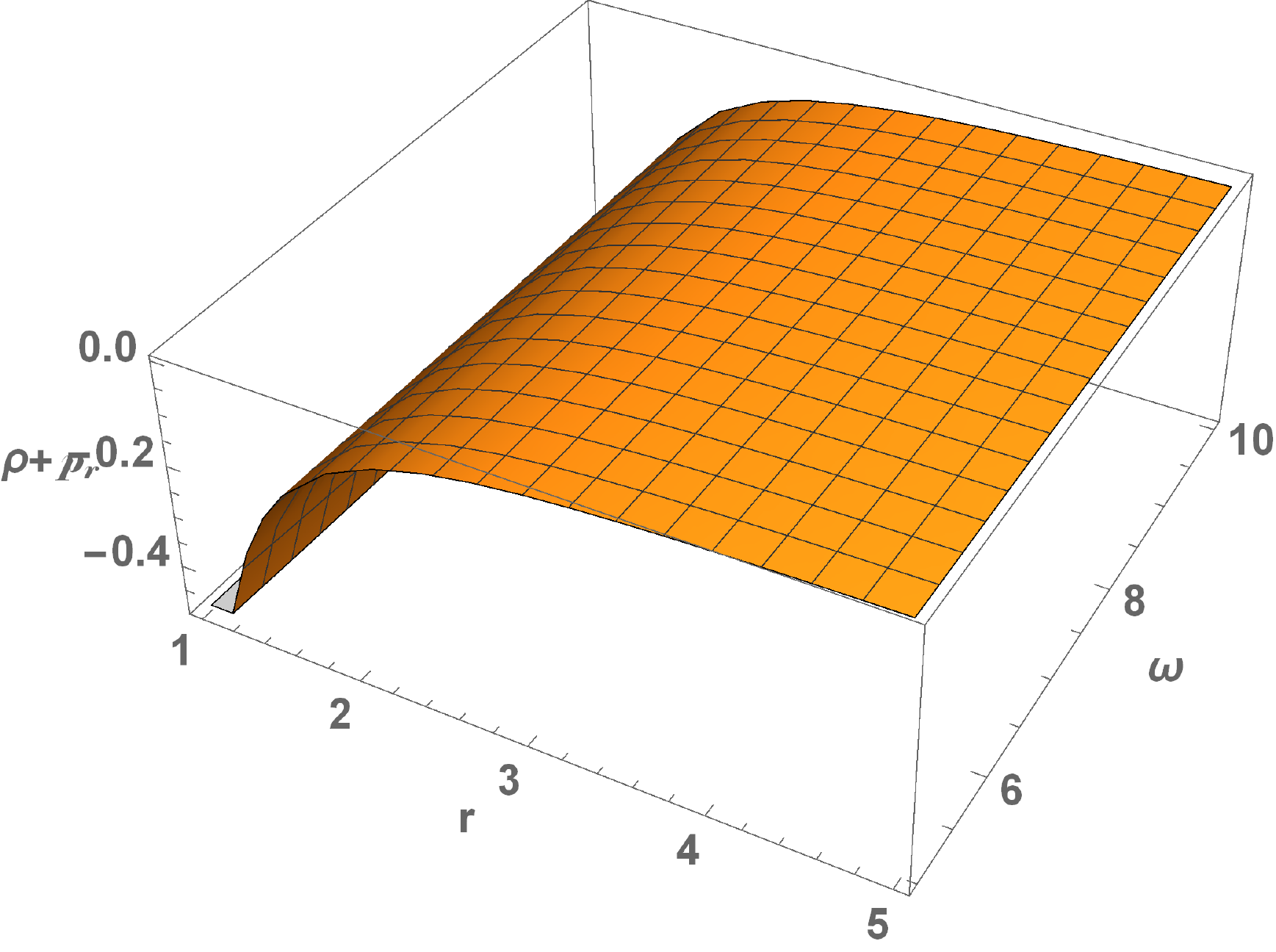}
\caption{\label{fig:7} Violation of NEC, $\rho+p_r\leq0$, with $b_0=1$ and $\lambda=-12$.}
\end{figure}

\begin{figure}[!h]
\includegraphics[scale=0.4]{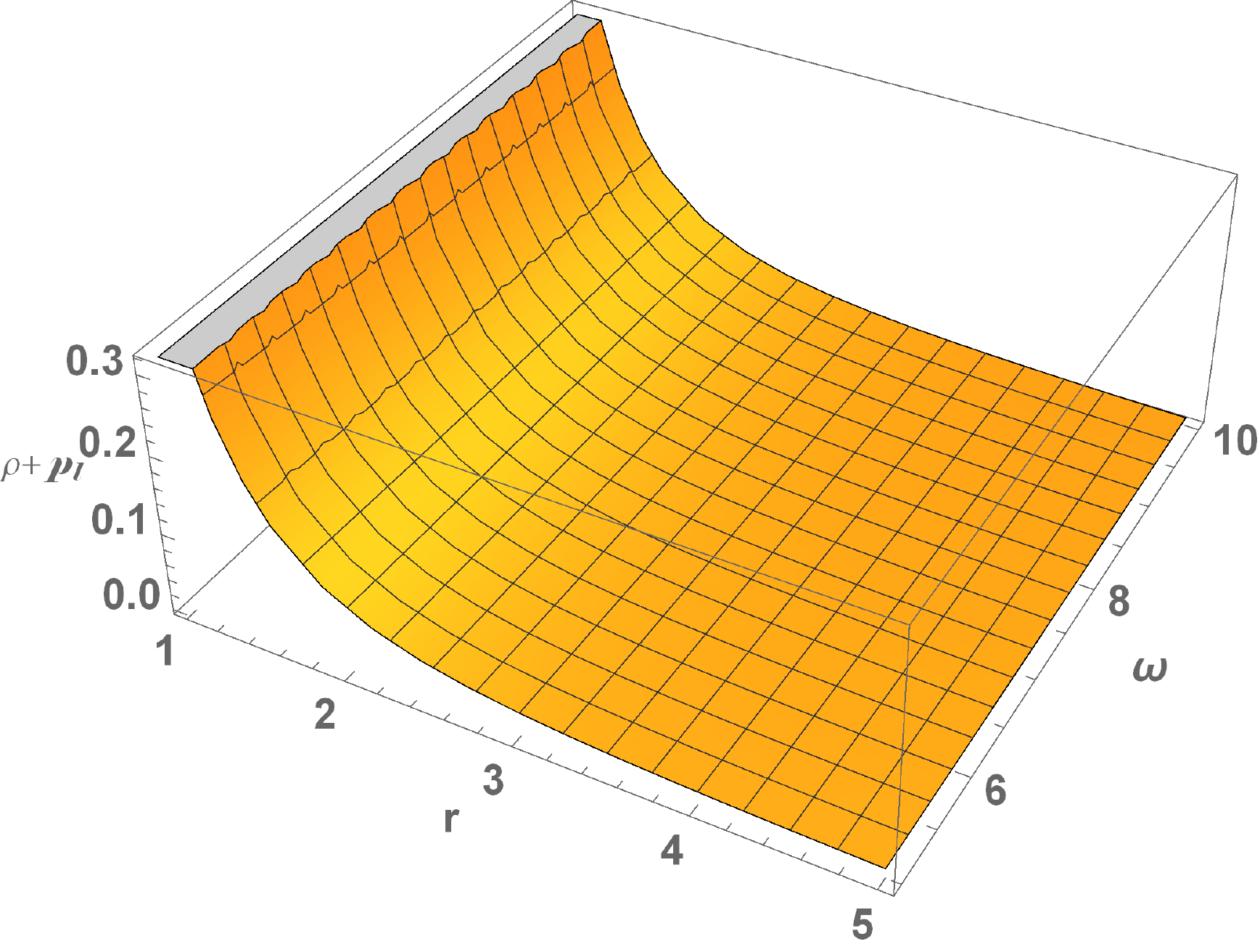}
\caption{\label{fig:8} Validity region of NEC, $\rho+p_l\geq0$, with $b_0=1$ and  $\lambda=-12$.}
\end{figure}

DEC for this model is

\begin{equation}\label{40}
\rho-p_r=\frac{b_0(1+\omega)r^{\frac{1-3\omega}{\omega}}}{\omega(8\pi+2\lambda)},
\end{equation}

\begin{equation}\label{41}
\rho-p_l=\frac{b_0(3-\omega)r^{\frac{1-3\omega}{\omega}}}{2\omega(8\pi+2\lambda)}.
\end{equation}
Those can be seen in Figs.9-10 below.

\begin{figure}[!h]
\includegraphics[scale=0.4]{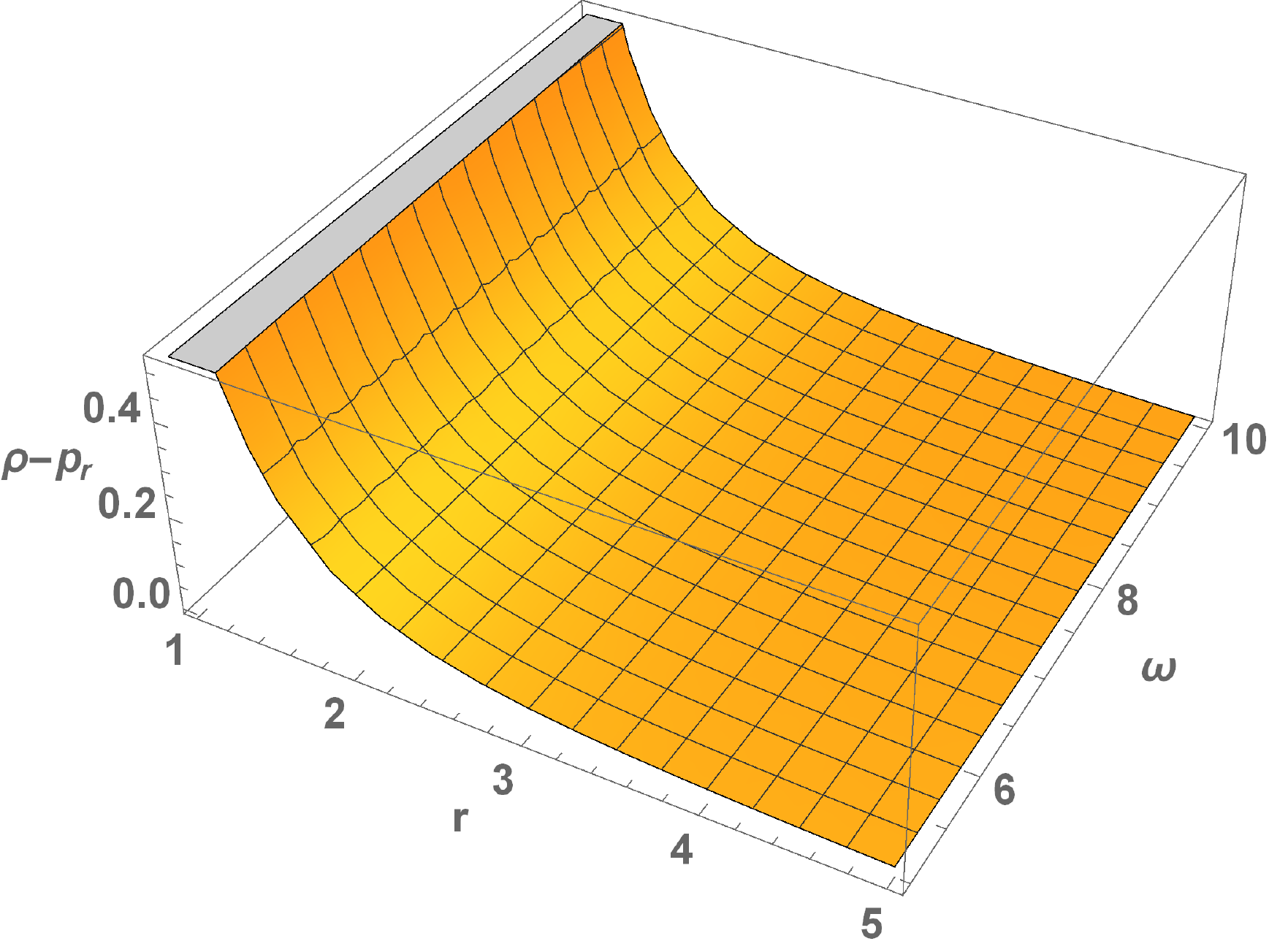}
\caption{\label{fig:9} Validity of DEC, $\rho\geq\vert p_r\vert$, with $b_0=1$ and  $\lambda=-12$.}
\end{figure}

\begin{figure}[!h]
\includegraphics[scale=0.4]{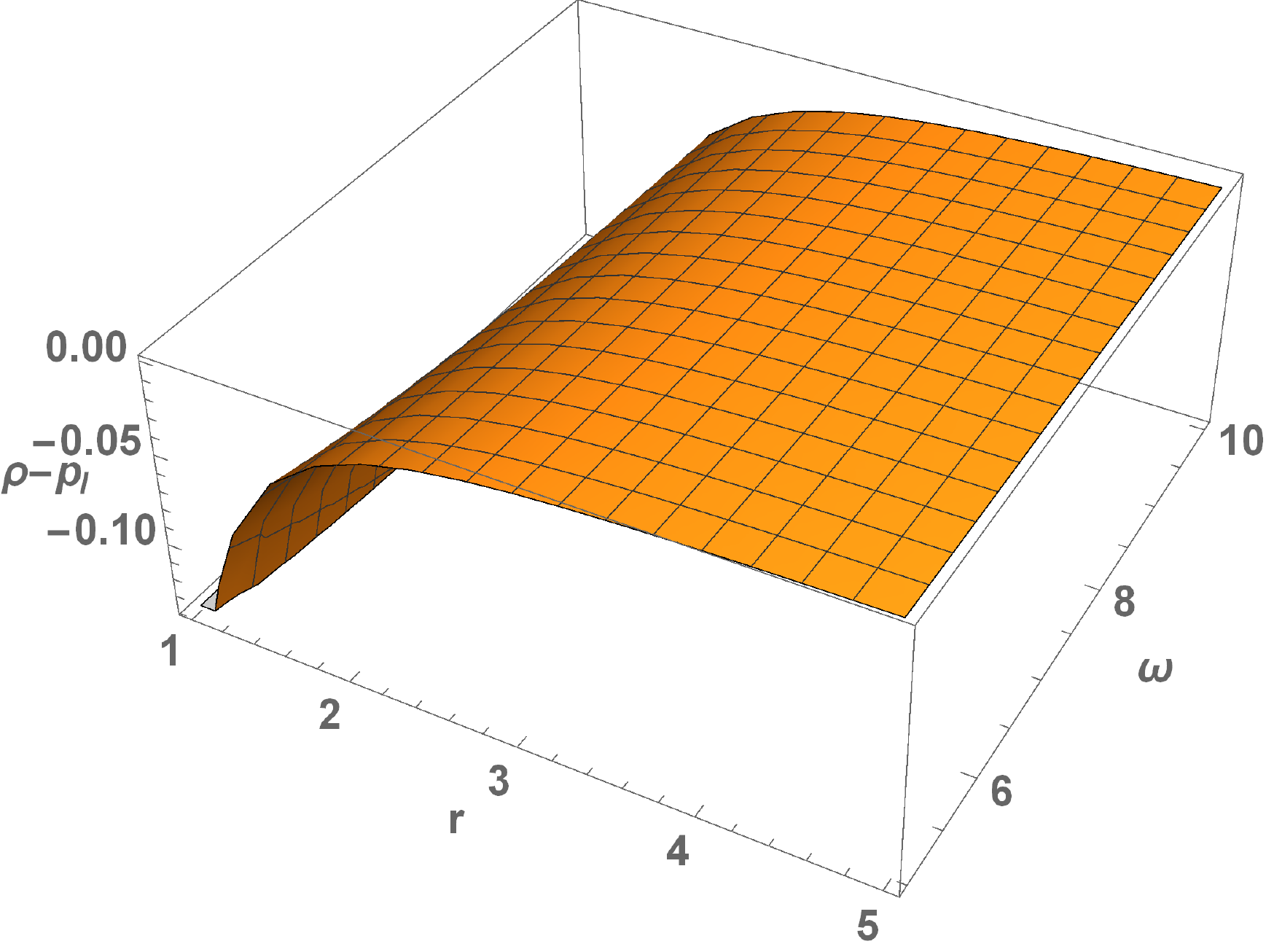}
\caption{\label{fig:10} Violation of DEC, $\rho\geq\vert p_l\vert$, with $b_0=1$ and  $\lambda=-12$.}
\end{figure}

\textbf{Case II - $\omega(r)=Br^m$}\\

Here, we consider $\omega(r)=Br^m$, where $B$ and $m$ are positive constants. From Eq.(32), the shape function now reads
\begin{equation}\label{42}
b=\exp\left(C-\frac{1}{Bmr^m}\right),
\end{equation}
where $C$ is a constant of integration.

At the throat of the WH we can obtain
\begin{equation}\label{43}
C=\ln r_0+\frac{1}{Bmr_0^m},
\end{equation}
so that
\begin{equation}\label{44}
b=\exp\left[\ln r_0+\frac{1}{Bm}\left(\frac{1}{r_0^m}-\frac{1}{r^m}\right)\right].
\end{equation}
Since $\omega>1$ and $m>0$, we have $r>r_0>\left(\frac{1}{B}\right)^{\frac{1}{m}}$, which satisfies the asymptotically flat condition for the metric. Hence the considered EoS is justified.

From Fig.11 below, one can note that when $r>r_0$, $b(r)-r<0$, which implies $\frac{b(r)}{r}<1$ when $r>r_0$. We can observe that $b(r)-r$ is a decreasing function of $r$ for $r \geq r_0$ and $b'(r_0)<1$, which satisfies the flaring out condition. Hence, the obtained shape function indeed represents a WH structure.

\begin{figure}[!h]
\includegraphics[scale=0.4]{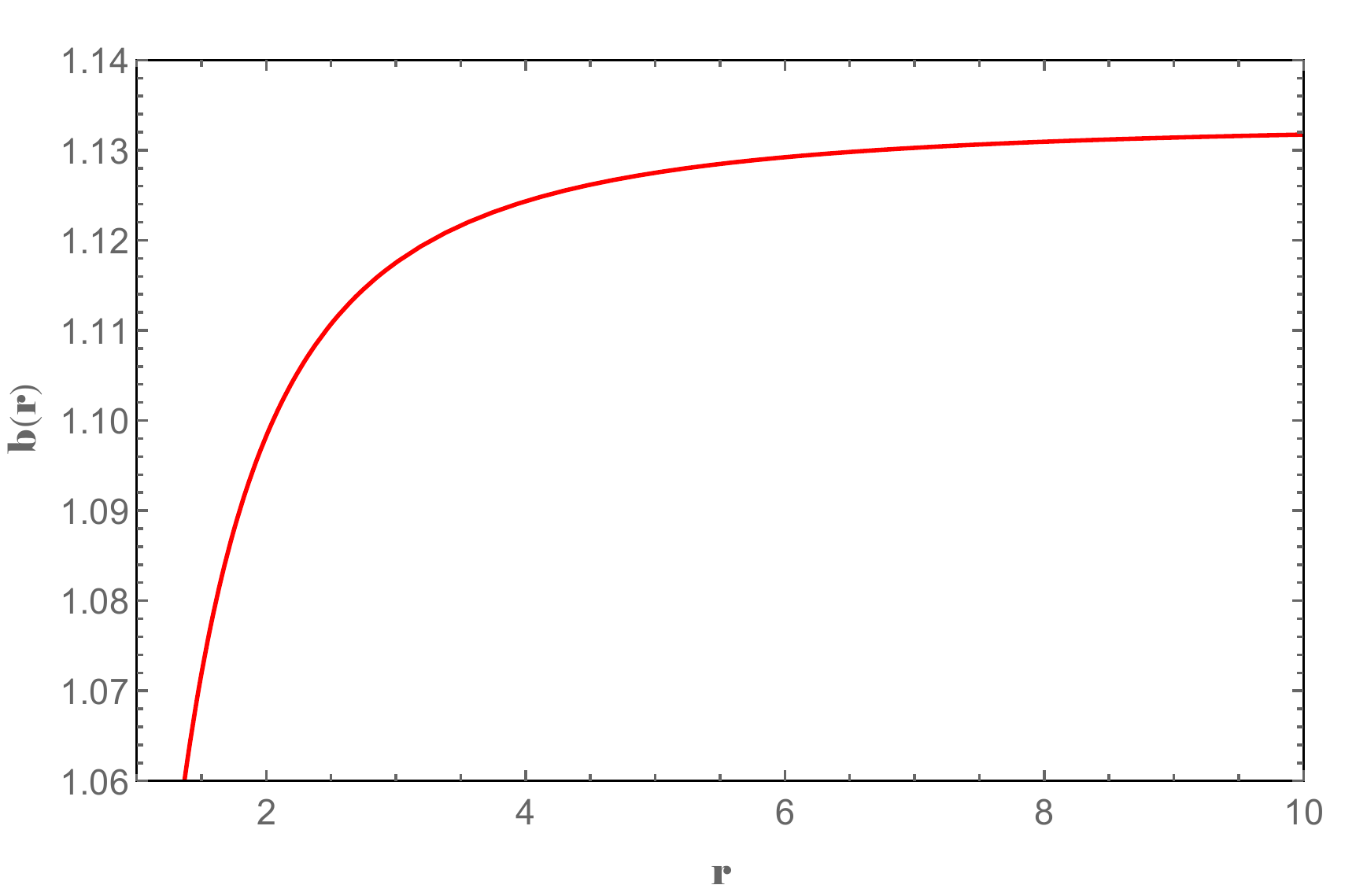}
\caption{\label{fig:11} $b(r)$ with $r_0=1, B=4$ and $m=2$.}
\end{figure}

Using Eq.(42) into Eqs.(18)-(20), we have
\begin{equation}\label{45}
\rho=\frac{\exp\left(C-\frac{1}{Bmr^m}\right)}{(8\pi+2\lambda)Br^{m+3}},
\end{equation}
\begin{equation}\label{46}
p_r=-\frac{\exp\left(C-\frac{1}{Bmr^m}\right)}{(8\pi+2\lambda)r^{3}},
\end{equation}
\begin{equation}\label{47}
p_l=\frac{(Br^{m+1}-1)\exp\left(C-\frac{1}{Bmr^m}\right)}{2(8\pi+2\lambda)Br^{m+3}}.
\end{equation}

NEC for this model reads
\begin{equation}\label{48}
\rho+p_r=\frac{(1-Br^{m})\exp\left(C-\frac{1}{Bmr^m}\right)}{(8\pi+2\lambda)Br^{m+3}},
\end{equation}
\begin{equation}\label{49}
\rho+p_l=\frac{(Br^{m+1}+1)\exp\left(C-\frac{1}{Bmr^m}\right)}{2(8\pi+2\lambda)Br^{m+3}}.
\end{equation}
Those are plotted in Figs.12-13 below.

\begin{figure}[!h]
\includegraphics[scale=0.4]{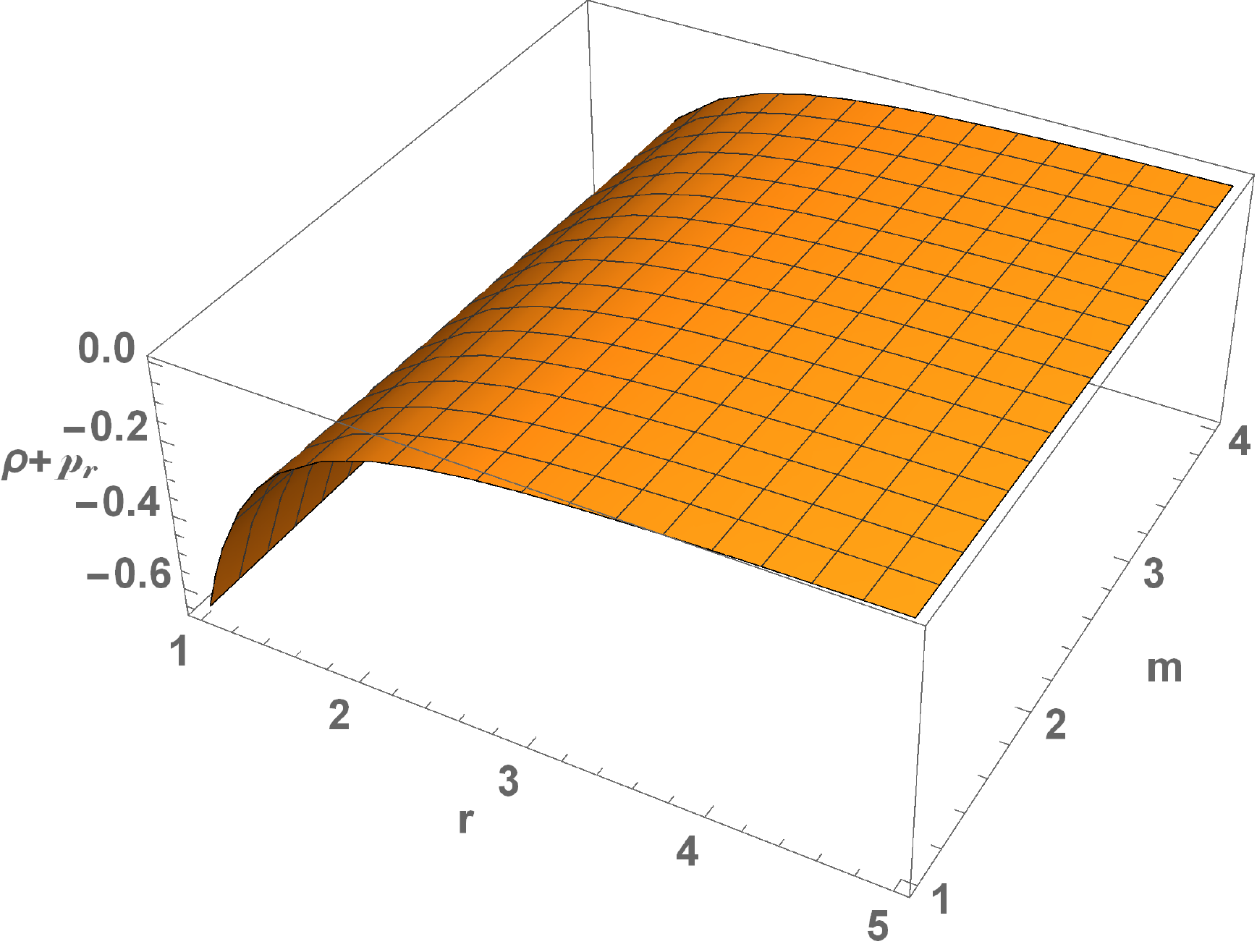}
\caption{\label{fig:12} Violation of NEC, $\rho+p_r\leq0$, with $r_0=1$, $B=4$ and $\lambda=-12$.}
\end{figure}
\begin{figure}[!h]
\includegraphics[scale=0.4]{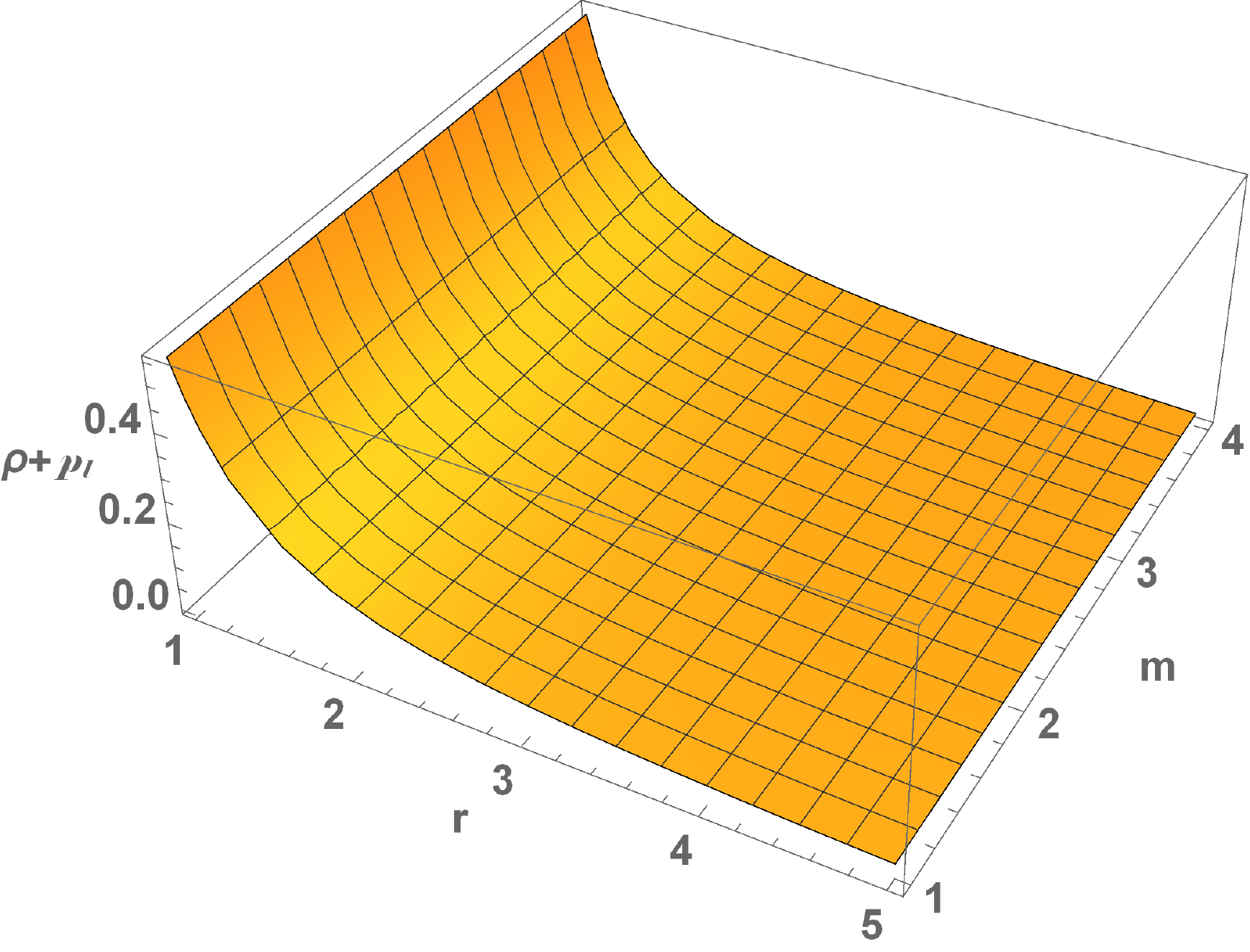}
\caption{\label{fig:13} NEC, $\rho+p_l\geq0$, with $r_0=1$, $B=4$ and $\lambda=-12$.}
\end{figure}

DEC for this model is
\begin{equation}\label{50}
\rho-p_r=\frac{(1+Br^{m})\exp\left(C-\frac{1}{Bmr^m}\right)}{(8\pi+2\lambda)Br^{m+3}},
\end{equation}
\begin{equation}\label{51}
\rho-p_l=\frac{(3-Br^{m+1})\exp\left(C-\frac{1}{Bmr^m}\right)}{2(8\pi+2\lambda)Br^{m+3}}.
\end{equation}
Those are plotted in Figs.14-15 below.

\begin{figure}[!h]
\includegraphics[scale=0.4]{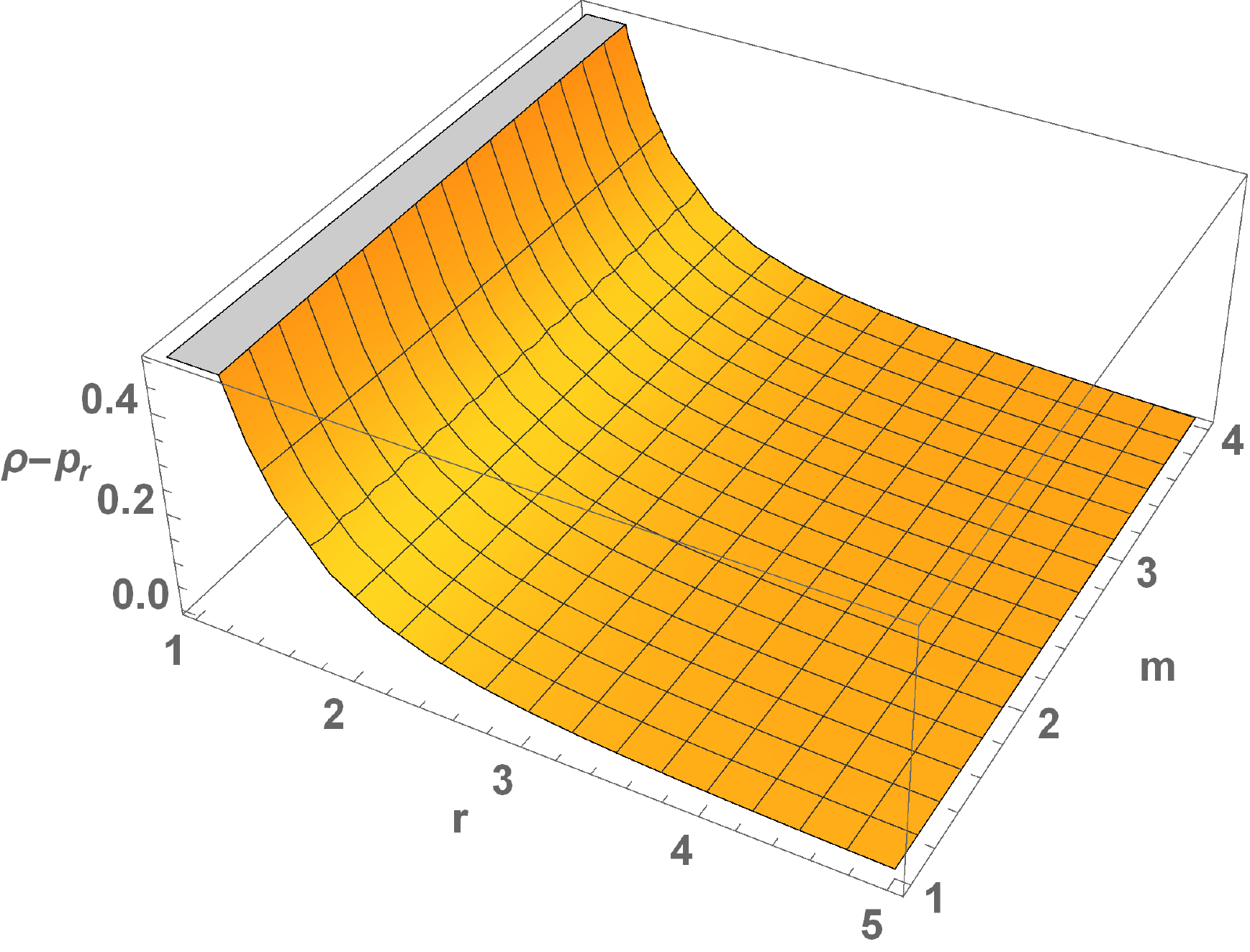}
\caption{\label{fig:14} Validity of DEC, $\rho\geq\vert p_r\vert$, with $r_0=1$, $B=4$ and $\lambda=-12$.}
\end{figure}
\begin{figure}[!h]
\includegraphics[scale=0.4]{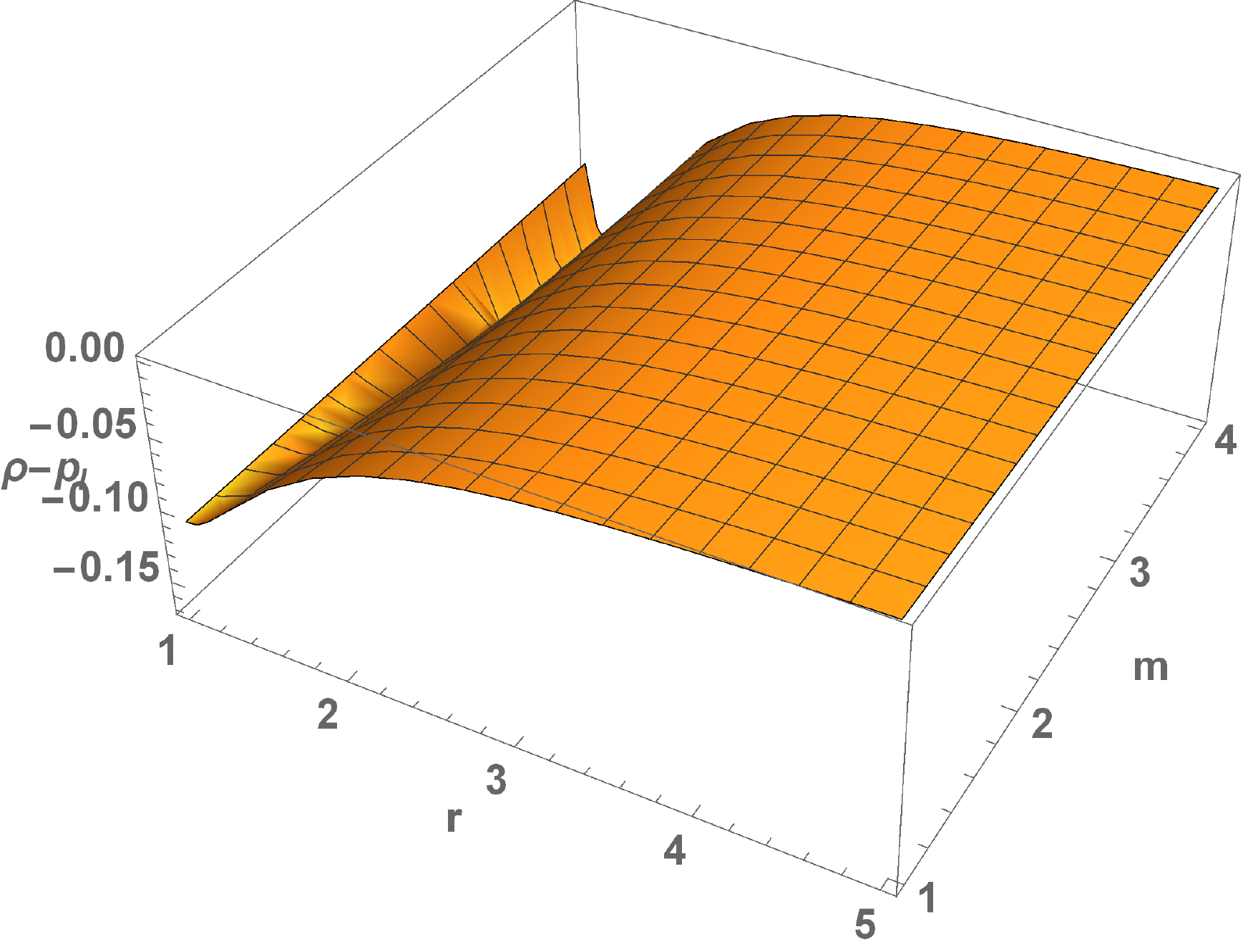}
\caption{\label{fig:15} Violation of DEC, $\rho\geq\vert p_l\vert$, with $r_0=1$, $B=4$ and $\lambda=-12$.}
\end{figure}

Moreover, for both Cases of Model 2, SEC yields $\rho+p_r+2p_l=0$.

\section{Discussion}

We have constructed, in the present article, different models of static WHs within the $f(R,T)$ theory of gravitation. In this section we will discuss the results we obtained for the geometrical and material content of those WHs.

As we have mentioned in the previous sections, the condition $b'(r_0)<1$, which is required to describe a WH solution, is satisfied in all the constructed models. For Model 1, $b'(r_0)=1+2n$, which is $<1$, since $n$ is negative, corresponding to Eqs.(22)-(23). For Case I of Model 2, $b'(r_0)=\frac{1}{\omega}<1$, since $\omega>1$ and for Case II, $b'(r_0)<1$ implies in $r_0>\left(\frac{1}{B}\right)^{\frac{1}{m}}$. 

Furthermore, we have assumed the redshift function is a constant ($U(r)=constant$), which implies that the tidal gravitational force experienced by a hypothetical traveller is null. 

In view of the seminal paper by M.S. Morris and K.S. Thorne \cite{morris/1988}, in such a reference, the authors have found that in a static WH, $\rho\sim r^{-2}$. We can see that our solution for $\rho$ in Case I of Model 2 predicts the same proportionality for $r$ when $\omega\rightarrow1$. Still in this case, our solution for $\rho$ agrees with those found for Morris-Thorne WHs with a cosmological constant \cite{lemos/2003} and WHs minimally violating NEC \cite{bouhmadi-lopez/2014}.

In General Relativity, the analysis of the geometry of traversable WHs has shown that NEC is violated at the WH throat \cite{morris/1988, visser/1995}. In our models, at the WH throat, the weaker inequality $\rho(r_0)+p_r(r_0)\leq0$ holds, which implies that NEC is violated. The authors in \cite{nandi/1998} obtained the validity of NEC, $\rho+p_r\geq0$, by considering negative energy density. SEC, on the other side, is satisfied in all of our models, i.e., $\rho+p_r+2p_l=0$, as one can check  Eqs.(18)-(20). 

It is interesting to remark that further analysis of $f(R,T)$ WHs may come from the implementation of higher order terms of $R$ and/or $T$ in the $f(R,T)$ function. Different forms for $f(R)$ have already been considered in WH analysis in \cite{bambi/2016,lobo/2009,duplessis/2015} for instance. Those can be added to functions of $T$ in order to get more generic $f(R,T)$ WHs.

Another valuable functional forms for the $f(R,T)$ function that can be used to construct WH models are those that predict a matter-geometry non-minimal coupling, such as $f(R,T)=R+\alpha RT$, which has already been applied in cosmological models, yielding observational acceptable results \cite{ms/2017}.

\acknowledgments PHRSM would like to thank S\~ao Paulo Research Foundation (FAPESP), grant 2015/08476-0, for financial support. PKS acknowledges the support of CERN, in Geneva, during an academic visit, where a part of this work was done.

\end{document}